\documentclass[aps,floats,floatfix,amssymb,prd,
onecolumn,superscriptaddress,nofootinbib,notitlepage,11pt]{revtex4-1}

\usepackage{slashed}
\usepackage{bm}
\usepackage{latexsym,amssymb,amsmath,float,url,mathrsfs}
\usepackage{latexsym}
\usepackage{graphicx}
\usepackage{epstopdf}
\usepackage{amsmath}
\usepackage{comment}
\usepackage{subfigure}
\usepackage{natbib}
\usepackage{hyperref}
\usepackage{pifont}
\usepackage{blindtext}
\usepackage{adjustbox}
\usepackage{multirow}
\usepackage{tabularx}
\usepackage[table]{xcolor}
\usepackage{orcidlink}
\usepackage{tikz}
\usetikzlibrary{tikzmark}
\usepackage{blkarray}
\usepackage{graphicx}
\usepackage{amsfonts}
\usepackage{soul}
\usepackage{amssymb}
\usepackage{amsmath}
\usepackage{cancel}
\usepackage{tcolorbox}

\usepackage{graphics,appendix,afterpage,makecell} 

\definecolor{oucrimsonred}{rgb}{0.6, 0.0, 0.0}
\definecolor{persianblue}{rgb}{0.11, 0.22, 0.73}
\definecolor{forestgreen}{rgb}{0.13,0.35,0.13}
\definecolor{lightgray}{rgb}{0.83, 0.83, 0.83}
 \hypersetup{colorlinks, citecolor=oucrimsonred, linkcolor=black, urlcolor=oucrimsonred}
\definecolor{cornellred}{rgb}{0.7, 0.11, 0.11}
\definecolor{navyblue}{rgb}{0.0, 0.0, 0.5}
\definecolor{amethyst}{rgb}{0.6, 0.4, 0.8}
\definecolor{yellow}{rgb}{1.0, 1.0, 0.0}
\definecolor{firebrick}{rgb}{0.7, 0.13, 0.13}
\definecolor{tangerineyellow}{rgb}{1.0, 0.8, 0.0}
\definecolor{deepfuchsia}{rgb}{0.76, 0.33, 0.76}
\definecolor{amber}{rgb}{1.0, 0.75, 0.0}
\definecolor{VioletRed4}{rgb}{0.55, 0.13, .32}
\definecolor{indiagreen}{rgb}{0.07, 0.53, 0.03}
\definecolor{VioletRed4}{rgb}{0.55, 0.13, .32}
\newcommand{\be}{\begin{equation}}
\newcommand{\ee}{\end{equation}}
\newcommand{\bea}{\begin{equation} \begin{aligned}}
\newcommand{\eea}{\end{aligned} \end{equation}}

\definecolor{oucrimsonred}{rgb}{0.6, 0.0, 0.0}
\newcommand\vertarrowbox[3][6ex]{%
  \begin{array}[t]{@{}c@{}} #2 \\
  \left\uparrow\vcenter{\hrule height #1}\right.\kern-\nulldelimiterspace\\
  \makebox[0pt]{\scriptsize#3}
  \end{array}%
}

\definecolor{verdechiaro}{rgb}{0.6,1,0.6}
\definecolor{giallochiaro}{rgb}{1,1,0.6}
\definecolor{bluscuro}{rgb}{0.15, 0.2, 0.9}
\definecolor{verdes}{rgb}{0.1, 0.5, 0.1}%
\definecolor{tangerineyellow}{rgb}{1.0, 0.8, 0.0}

\definecolor{americanrose}{rgb}{1.0, 0.01, 0.24}
\definecolor{cobalt}{rgb}{0.0, 0.28, 0.67}
\definecolor{brandeisblue}{rgb}{0.0, 0.44, 1.0}
\definecolor{mycolor}{rgb}{0.0, 0.0, 0.5}%navyblue
\definecolor{oxfordblue}{rgb}{0.0, 0.13, 0.28}
\definecolor{azure}{rgb}{0.0, 0.5, 1.0}
\definecolor{turquoiseblue}{rgb}{0.0, 1.0, 0.94}
\newtcolorbox{mynewbox}[1]{colback=white!5!white,colframe=azure!75!black,fonttitle=\bfseries,title=#1}
\newtcolorbox{mybox}{colback=mycolor!5!white,colframe=azure!75!black}
\newtcolorbox{mynamedbox}[1]{colback=mycolor!5!white,colframe=azure!75!black,title=#1}
\definecolor{venetianred}{rgb}{0.78, 0.03, 0.08}
\newtcolorbox{mynamedbox1}[1]{colback=venetianred!5!white,colframe=venetianred!80!black,title=#1}
\newtcolorbox{mynamedbox2}[1]{colback=azure!5!white,colframe=azure!80!black,title=#1}

\definecolor{verdes}{rgb}{0.1, 0.5, 0.1}%
\definecolor{cornellred}{rgb}{0.7, 0.11, 0.11}

\definecolor{VioletRed4}{rgb}{0.55, 0.13, .32}

\hypersetup{
     colorlinks   = true,
     citecolor    = violet,
     urlcolor     = violet,
     linkcolor    = violet}

\definecolor{rossocorsa}{rgb}{0.83, 0.0, 0.0}

\usepackage[normalem]{ulem}

\begin{document}
%%%%%%%%%%%%%%%%%%%%%%%%%%%%%%%%%%%%%%%%%%%%%%%%%%%%%%%%%%%  FRONT PAGE

\title[]{A Short Note on the Love Number \\ of Extremal  Reissner-Nordstr\o m  and Kerr-Newman Black Holes}

\author{Alex Kehagias\orcidlink{}}
\affiliation{Physics Division, National Technical University of Athens, Athens, 15780, Greece}

\author{Davide Perrone\orcidlink{0000-0003-4430-4914}}
\affiliation{Department of Theoretical Physics and Gravitational Wave Science Center,  \\
24 quai E. Ansermet, CH-1211 Geneva 4, Switzerland}

\author{Antonio Riotto\orcidlink{0000-0001-6948-0856}}
\affiliation{Department of Theoretical Physics and Gravitational Wave Science Center,  \\
24 quai E. Ansermet, CH-1211 Geneva 4, Switzerland}

%%%%%%%%%%%%%%%%%%%%%%%%%%%%%%%%%%%%%%%%%%%%%%%%%%%%%%%%%%%%%%%%%%%%%

\begin{abstract}
\noindent
We provide a simple proof of why the Love number vanishes for  
extremal Reissner-Nordstr\o m  and Kerr-Newman black holes. The argument is based on a conformal inversion isometry of the spacetime connecting the horizon with large distances.

\end{abstract}

%%%%%%%%%%%%%%%%%%%%%%%%%%%%%%%%%%%%%%%%%%%%%%%%%%%%%%%%%%%%%%%%%%%
\maketitle

%%%%%%%%%%%%%%%%%%%%%%%%%%%%%%%%%%%%%%%%%%%
%\noindent\textit{Introduction.} 
\section*{1. Introduction}
%%%%%%%%%%%%%%%%%%%%%%%%%%%%%%%%%%%%%%%%%%%
\noindent
General relativity predicts the existence of gravitational waves and black holes which  are both tested in current experiments detecting  the gravitational waves produced by black hole mergers~\cite{LIGOScientific:2021sio}. In particular, 
when the  orbital separation  of two compact objects  is small enough,  the tidal effects  between the two bodies become relevant and are usually described in terms of the so-called tidal Love numbers~\cite{poisson_will_2014}. The tidal Love numbers depend on the internal properties and structure of the deformed compact object and impact the gravitational wave emission at fifth post-Newtonian order~\cite{Flanagan:2007ix}. 

Interestingly enough, the  tidal Love numbers of non-rotating and spinning black holes have been   found to be exactly zero~\cite{Binnington:2009bb,Damour:2009vw,Damour:2009va,Pani:2015hfa,Pani:2015nua,Gurlebeck:2015xpa,Porto:2016zng,LeTiec:2020spy, Chia:2020yla,LeTiec:2020bos},  revealing underlying hidden symmetries of General Relativity~\cite{Hui:2020xxx,Charalambous:2021mea,Charalambous:2021kcz,Hui:2021vcv,Hui:2022vbh,Charalambous:2022rre,Ivanov:2022qqt,Katagiri:2022vyz, Bonelli:2021uvf,Kehagias:2022ndy,BenAchour:2022uqo,Berens:2022ebl,DeLuca:2023mio, Rai:2024lho,Riva:2023rcm}, at least at the linear level. Having zero Love number for the Schwarzschild and Kerr black holes is relevant to distinguish them from neutron stars in sub-solar mergers and eventually investigating the primordial nature of the black holes \cite{Crescimbeni:2024cwh,Riotto:2024ayo}.

A simple and direct way to test the vanishing (or not) of the Love number is to use a  massless scalar field $\phi$ (a proxy for the helicity two tensor degrees of freedom) and solve its dynamics in the static case. At infinity, for asymptotically flat spacetimes, the general solution is of the type

\be
\label{a}
\phi(r\rightarrow \infty)= a_\ell \,r^\ell+\frac{b_\ell}{r^{\ell+1}}.
\ee
The  growing mode $\sim r^\ell$ represents the tidal force by which the reaction of the spacetime metric generated by the body (in our case black hole) is tested. The reaction is measured by the decaying mode $\sim r^{-\ell-1}$. 
The Love number is fixed by the ratio $b_\ell/a_\ell$ and vanishes if $b_\ell=0$. For non-rotating and spinning black holes this happens because  the decaying mode $r^{-\ell-1}$ is mapped into the divergent solution at the horizon and therefore must be excluded from the set of physical solutions. This connection is best seen through a ladder symmetry which appears in the equation of motion of the scalar field  in the static case \cite{Hui:2021vcv}. The origin of the ladder symmetry can be also understood by the fact that, in the static limit, the massless scalar field effectively propagates
in a AdS$_2\times$S$_2$ spacetime and the latter symmetry is part of its conformal isometries 
\cite{Hui:2022vbh}.

In this paper we make use of  a  symmetry argument to show that  the Love number of extremal Reissner-Nordstr\o m  and Kerr-Newman black holes vanishes. The argument is astonishingly simple. The extremal Reissner-Nordstr\o m  and Kerr-Newman spacetimes with metric $g_{\mu\nu}$ possess a conformal inversion isometry connecting the near horizon region to the large distance region

\begin{equation}
  r\rightarrow \frac{1}{r},\,\, g_{\mu\nu}\rightarrow\Omega^2\,g_{\mu\nu},\,\,\Omega^2\sim \frac{1}{r^2}.  
\end{equation}
Since such spacetimes are Ricci flat, the  massless scalar field action is invariant under such a inversion transformation if the field is properly transformed, 
\begin{equation}
  \phi(r)\rightarrow \frac{1}{r}\phi(1/r).  
\end{equation}
In the static limit, the equation of motion of the scalar field reduces to that of massless scalar field propagating in flat Minkowski where no tidal reactions may be generated. Indeed,  the solution  for the scalar field dynamics is 
that of Eq. (\ref{a}) in all the space (and not only at large distances), which manifestly satisfies the inversion property. Requiring that the solution is finite at the horizon sets $b_\ell=0$ and the Love number is correspondingly zero. 
This symmetry arguments differs from the one of the ladder symmetry since the latter is a good symmetry only in the near-zone (frequencies much smaller than the inverse distance) limit \cite{Hui:2022vbh}, while the conformal inversion symmetry is an isometry of the original spacetime. 

\vskip 0.5cm
%%%%%%%%%%%%%%%%%%%%%%%%%%%%%%%%%%%%%%%%%%%
%\noindent\textit{Some basic concepts.} 
\section*{2. The  Extremal Reissner-Nordstr\o m Black Hole }
\noindent
The metric of the extreme Reissner-Nordstr\o m black hole reads

\begin{equation}
{\rm d}s^2=-\left(1-\frac{m}{\widetilde r}\right)^2{\rm d}t^2+\left(1-\frac{m}{\widetilde r}\right)^{-2}{\rm d}\widetilde r^2+\widetilde r^2{\rm d}\Omega^2.
\end{equation}
By setting  $r=(\widetilde r-m)$, we can rewrite it in a more convenient form in the so-called isotropic coordinates

\begin{equation}
{\rm d}s^2=-\left(1+\frac{m}{r}\right)^{-2}{\rm d}t^2+\left(1+\frac{m}{r}\right)^{2}\left({\rm d}r^2+r^2{\rm d}\Omega^2\right),
\end{equation}
where now the position of the horizon is at $r=0$. Key to our discussion is the fact that the extremal Reissner-Nordstr\o m metric  $g_{\text{\tiny ERN}}$ admits a discrete conformal isometry
 (i.e. isometry up to multiplication by a conformal factor) generated by the spatial inversion defined by the change of the coordinate \cite{CT}

\begin{equation}
r\rightarrow \frac{m^2}{r}, \,\,\,\,
g_{\text{\tiny ERN}}\rightarrow \Omega^2 g_{\text{\tiny ERN}},\,\,\,\,\Omega=\frac{m}{r},
\end{equation}
which exchanges the horizon at $r=0 $ and infinity. 

%Indeed, one  readily finds 

%\begin{equation}
%{\rm d}s^2=\frac{m^2}{y^2}\left[-\left(1+\frac{m}{y}\right)^{-2}{\rm d}t^2+\left(1+\frac{m}{y}\right)^{2}\left({\rm d}y^2+y^2{\rm d}\Omega^2\right)\right],
%\end{equation}
%which is  the extremal Reissner-Nordstr\o m metric in isotropic coordinates, up to the conformal prefactor $m^2/y^2$.

Consider now the action of a massless scalar field $\phi$ 
in the background of the Reissner-Nordstr\o m black hole. Since the Reissner-Nordstr\o m spacetime is Ricci flat, the action of such massless scalar field is invariant if, besides the conformal inversion of the metric,  we also transform the field as~\cite{Bizon:2012we}

\begin{equation}
  \phi(t,r)\rightarrow \frac{m^2}{r}\phi\left(t,\frac{m^2}{r}\right).  
\end{equation}
In other words, if $\phi(t,r)$ is a solution of the equation of motion, then also $\phi(t,m^2/r)/r$ is.  This can be explicitly seen once it is decomposed in spherical harmonics and the dependence on time is encoded in the frequency $\exp(-i\omega t)$

\begin{equation}
     \left(1+\frac{m}{r}\right)^4\omega^2 \phi_\ell + \frac{2}{r } \partial_r \phi_\ell + \partial_r^2 \phi_\ell - \frac{\ell(\ell+1)}{r^2}\phi_\ell =0.
\end{equation}
Close to the horizon the solutions are combinations of the Bessel functions

\begin{equation}
\phi_{0\ell}(r,\omega)= A_\ell  \sqrt{\frac{m^2 \omega}{ r}}J_{ (\ell + 1/2)}\left( \frac{ m^2 \omega }{r} \right)+B_\ell \sqrt{\frac{m^2\omega} {r}}J_{ -(\ell + 1/2)}\left( \frac{ m^2\omega }{r}\right).
\end{equation}
At large radii the solution is expressed in terms of the spherical Bessel functions

\begin{equation}
    \phi_{\infty\ell}(r,\omega)= C_\ell \, j_{\ell} (\omega r) + D_\ell \,y_{\ell} (\omega r).
\end{equation}
Recalling now that $j_\nu(x)=\sqrt{\pi/2 x}J_{\nu+1/2}(x)$, one can recognize that $\phi_{\infty\,\ell}(t,r)=\phi_{0\,\ell}(t,m^2/r)/r$ by choosing appropriately the constants of integration.
This makes the conformal inversion isometry a symmetry of the scalar action manifest.

The static case, appropriate to look at the static Love number, corresponding to $\omega=0$ is even simpler and striking. In such a case, the equation reduces to that of a massless scalar field propagating in an effective Minkowski spacetime and  the exact solution of the equation of motion is (in all the space)

\begin{equation}
  \phi_\ell(r)=a_\ell\, r^\ell+\frac{b_\ell }{r^{\ell+1}}.
\end{equation}
Clearly, the conformal inversion operation onto the scalar field maps one solution into the other. 

The static Love number can now be readily deduced to be zero. Indeed of the two solutions one has to retain only the solution going like $r^\ell$   as it is the only regular at the horizon $r=0$ and also plays the role of the tidal force. At the same time the regularity at the horizon imposes $b_\ell=0$ and therefore the static Love number vanishes. This can be understood also by noting that in the static case the massless scalar field propagates in a spacetime which is effectively a flat Minkowski spacetime and therefore no reaction to tidal forces appear.
This simple derivation is in agreement with the statement that the Love number of the 
Reissner-Nordstr\o m black hole is zero \cite{Rai:2024lho}.

\section*{2. The  Extremal Kerr-Newman Black Hole }
\noindent
In the case of an  extremal Kerr-Newman black hole, an axially symmetric massless scalar field  does have the same conformal inversion symmetry \cite{Bizon:2012we}. Indeed,  the extreme Kerr-Newman metric $g_{\text{\tiny KN}}$ written in terms of isotropic Boyer-Lindquist coordinates (so that the horizon is located at $r=0$) reads

\begin{eqnarray}
 {\rm d}s^2&=&-\frac{\rho^2r^2}{A}  {\rm d}t^2+\frac{A\sin^2\theta}{\rho^2}\left({\rm d}\varphi-\omega{\rm d}t\right)^2+\frac{\rho^2}{r^2}{\rm d}r^2+\rho^2{\rm d}\theta^2,\nonumber\\
 \rho^2&=&(r+m)^2+a^2\cos^2\theta,\,\, A=\left[(r+m)^2+a^2\right]^2-a^2 r^2\sin^2\theta,\,\,\omega=\frac{m^2+a^2+2m r}{A}a.
\end{eqnarray}
The equation of motion for an axially symmetric massless scalar field satisfies the equation

\begin{equation}
\label{eqKN}
-\frac{A}{r^2}\partial_t^2\phi+\partial_r\left(r^2\partial_r\phi\right)+\frac{1}{\sin\theta}\partial_\theta\left(\sin\theta\partial_\theta\phi\right)=0.    
\end{equation}
A direct computation shows that if $\phi(t,r,\theta)$ is a solution, then also $\phi(t,(m^2+a^2)/r,\theta)\,/r$ is a solution,  under the conformal inversion
\begin{equation}
r\rightarrow \frac{m^2+a^2}{r}
\end{equation}
which exchanges the horizon at $r=0 $ and infinity.
In the static case the solution of  Eq. (\ref{eqKN}) reads

\begin{equation}
\phi(r,\theta)=\sum_{\ell=0}^\infty\left[a_\ell r^\ell+\frac{b_\ell }{r^{\ell+1}}\right]P_\ell(\cos\theta)
\end{equation}
and one concludes that the static Love number must be zero as the only solution regular at the horizon is the one with $b_\ell=0$. Again, another way to understand the result is that the action of the massless degree of freedom in the static case is equivalent to that of a massless scalar in  flat spacetime. 

Our findings show once more the relevance of underlying symmetries to uncover the physics of black holes.

%%%%%%%%%%%%%%%%%%%%%%%%%%%%%%%%%%%%%%%%%%%
\noindent

%%%%%%%%%%%%%%%%%%%%%%%%%%%%%%%%%%%%%%%%%%%

\vskip 0.5cm
%%%%%%%%%%%%%%%%%%%%%%%%%%%%%%%%%%%%%%%%%%%

%\newpage
\begin{acknowledgments}
\noindent
 We thank V. De Luca for useful comments. A.R.  acknowledges support from the  Swiss National Science Foundation (project number CRSII5\_213497) and from 
the Boninchi Foundation for the project ``PBHs in the Era of GW Astronomy''. The work of D. P. is supported by the Swiss National Science Foundation under grants no. 200021- 205016 and PP00P2-206149.
\end{acknowledgments}

\bibliography{Draft}
\end{document}